# Superconducting MgB$_2$ thin films by pulsed laser deposition.

S. R. Shinde, S. B. Ogale, R. L. Greene and T. Venkatesan
*Center for Superconductivity Research, Department of Physics, University of Maryland, College Park, MD 20742-4111.*

P. C. Canfield, S. L. Bud'ko, G. Lapertot, C. Petrovic
*Ames Laboratory, U. S. Department of Energy and Department of Physics and Astronomy, Iowa State University, Ames, Iowa 50011.*

Growth of MgB$_2$ thin films by pulsed laser deposition is examined under *ex situ* and *in situ* processing conditions. For the *ex situ* process, Boron films grown by PLD were annealed at 900 C with excess Mg. For the *in situ* process, different approaches involving ablation from a stoichiometric target under different growth conditions, as well as multilayer deposition involving interposed Mg layers were examined and analyzed. Magnetic measurements on *ex situ* processed films show T$_C$ of ~39 K, while the current best *in situ* films show a susceptibility transition at ~ 22 K.

The discovery of superconductivity in MgB$_2$ with T$_c$ ~ 39 K by Akimitsu and coworkers has caused a tremendous excitement in the scientific community due to the rather simple binary intermetallic character of the material system, and the prospects for further discoveries of materials with even higher T$_c$ in this whole new class of compounds.[1] Quickly following the initial discovery, Bud'ko et al.[2], Finnemore et al.[3], and Canfield et al.[4], established some key facts pertaining to Boron isotope effect as well as the thermodynamic and transport properties of this system, and demonstrated a method to fabricate dense wires of this new superconductor. The T$_c$, H$_{c2}$(T) and J$_c$ data on these dense wires showed a better performance as compared to their sintered pellet form. The development of wire is undoubtedly a key to the potential technological utilization of the new superconductor. Another development which is very much needed for realization of a range of high-tech applications of MgB$_2$ superconductor involving Josephson junctions and hetero-structure based devices is the reproducible preparation of epitaxial thin films of this material, preferably grown by an *in situ* process. There has been a report of MgB$_2$ film prepared by a possible post-annealing process & pulsed laser deposition.[5] In this letter, we address this issue and present the results of our experiments on the growth of MgB$_2$ films by the pulsed laser deposition (PLD) technique, either as an entirely *in situ* process, or as an *ex situ* process involving PLD of pure Boron followed by annealing in Mg vapor.

For the case of pulsed laser deposition with *ex situ* annealing in Mg, we deposited Boron films by pulsed laser deposition and reacted these with Mg in a sealed Ta tube in Mg vapor at 900 C. The Boron films were deposited at 800 C on SrTiO$_3$ (100) and (111) substrates at the rate of 0.035 Å/pulse, the thickness being ~1000 Å. We also compared these results with films deposited from an MgB$_2$ target onto STO substrates and subjected to Mg vapor reaction. For the case of *in situ* deposition, three different approaches were attempted: (a) PLD from a MgB$_2$ sintered target, (b) PLD of multi-layers of MgB$_2$ and Mg followed by *in situ* annealing at high temperature, and (c) PLD of multilayers of Mg and B followed by *in situ* annealing at high temperature. Different cap layers at the bottom as well as top of the film sandwich were examined to control interface reaction and to avoid Mg escape during the high temperature anneal process. The details of these optimization studies will be reported separately in a longer paper. Here we present the conditions and results for the case which yielded highest T$_c$ by an *in situ* process within the parameter framework we have examined thus far. For this case, STO (111) was used as the substrate. First a 10 nm thin layer from MgB$_2$ target was grown at 500 C. This was followed by the deposition of multilayers from MgB$_2$ and Mg targets at 350 C with nominal thicknesses of a few nanometers each, with 400 bilayers in a sandwich. Finally, a cap layer of ~ 10 nm was deposited from MgB$_2$ target. Such a configuration was annealed in situ at 900 C for 30 min. The heating rate was 40 C/min and the cooling rate was 5 C/min.

In Fig. 1 is shown the x-ray diffraction pattern for the film grown on STO (001) substrate by the *ex situ* process. Only one sharp line corresponding to the desired MgB$_2$ phase is seen indicating highly oriented, if not epitaxial, character of the film. The other two small peaks correspond to Mg. In fact, small Mg balls were found to be sticking on the film surface after the *ex situ* reaction in Mg vapor. These could not be removed by wiping with a cotton stub, but could be scrapped with a blade.

These could also be removed by low temperature evaporation. We are exploring ways to remove this excess Mg by evaporation or by selective etching, without affecting the underlying film properties.

In Fig. 2 (a) we show the zero field cooled (ZFC) magnetization as a function of temperature for the *ex situ* processed film for two magnetic field values of 25 Oe and 200 Oe. A fairly sharp transition is seen near 39 K with a broad tail, for the field of 25 Oe. The transition shifts to lower temperature and broadens with an increase in field to 200 Oe, as expected. However, if one starts with a film deposited from a $MgB_2$ target, even after post annealing in Mg a well defined unique $T_c$ is not seen. Instead a second transition at about 20 K is seen (Fig. 2 (b) and 2 (c) for films on STO(100) and STO(111) substrates, respectively).

In the case of the *in situ* processed films, the transition as obtained from magnetization for the best case (conditions described earlier) is shown in Fig. 3 (a). It shows a transition at 22 K with a tail. These lower values of $T_c$ could be due to departure from perfect stoichiometry in view of the possible variability of individual thicknesses in the multilayer, and also possibly loss of Mg at high temperature. We believe that careful thickness calibration followed by further optimization of the annealing sequence with an added suitably chosen cap layer would enhance the superconducting transition temperature as well as its quality. On films deposited directly from $MgB_2$ target at 800 C, a sharp transition at ~ 8K has been seen (Fig. 3(b)). For comparison, the ac susceptibility data of $MgB_2$ film prepared by *ex situ* process is shown in Fig. 3(c).

The observation of superconducting transitions at 8K, 22K and 39K in films under different processing conditions is noteworthy. Specifically, the transition near 20-22 K was observed in the *ex situ* process when implemented on films grown from $MgB_2$ target on STO(100) and STO(111) substrates and then magnesiated, as well as in the *in situ* films. These 8 K and 22 K transitions may be due to stabilization of different phases of $MgB_y$ depending upon Mg deficiency. The sharpness of the transitions at lower temperatures raises an interesting question from the physics standpoint: do these transitions suggest occurrence of superconductivity in other stoichimetric phases of Mg-B system such as $MgB_4$ and $MgB_6$ or other, yet to be determined, $MgB_y$ compound?

In summary, *ex situ* and *in situ* processes are examined for the growth of superconducting $MgB_2$ films on $SrTiO_3$ substrate. The *ex situ* process is shown to yield a $T_c$ of ~39 K, while the *in situ* process optimized thus far yields a $T_c$ of ~ 22 K. In the *ex situ* process, the highest $T_c$ is obtained by magnesiating pure Boron films, rather than films made from $MgB_2$ target. This suggests that a careful study of reaction and diffusion kinetics as a function of temperature and role of intermediate phases is warranted in this system. In addition, it demonstrates that if components can be coated with Boron films, then exposure to Mg vapor can transform these films into $MgB_2$. For the *in situ* process, maintaining correct stoichiometry, especially retaining Mg in the film is a critically important issue yet to be fully resolved. Surface morphology of the growing film as influenced by the wetting properties of Mg films as a function of deposition temperature is an important issue, which deserves attention. While we have succeeded in growing films with $T_c$ close to bulk value by the *ex situ* process, an *in situ* process would be needed for realization of successful junction development. In the mean time, we are studying the systematic of the dependence of $J_c$ on temperature and field on the high quality *ex situ* films.

The authors would like to acknowledge help from R.P. Sharma, A. Godinez, and M. Lewis during experiments. The work at University of Maryland is supported under ONR grant no. ONR-N000149611026 and NSF-MERSEC program under grant no. DMR0080008. Ames Laboratory is operated for the U. S. Department of Energy by Iowa State University under contract no. W-7405-ENG-82 and the work at Ames Laboratory is supported by the Director of Energy Research, Office of Basic Energy Sciences.

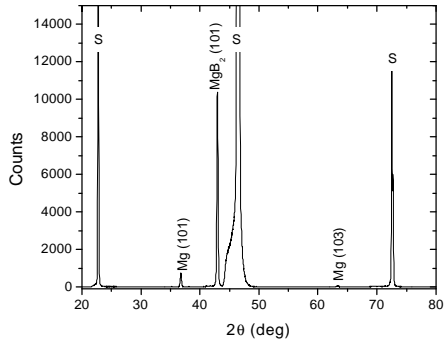

Fig.1: X-ray diffraction pattern of MgB$_2$ film prepared by *ex-situ* process on STO (100). The peaks denoted by "S" are the substrate peaks.

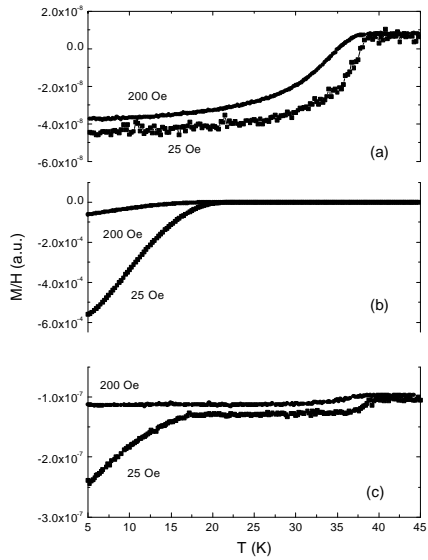

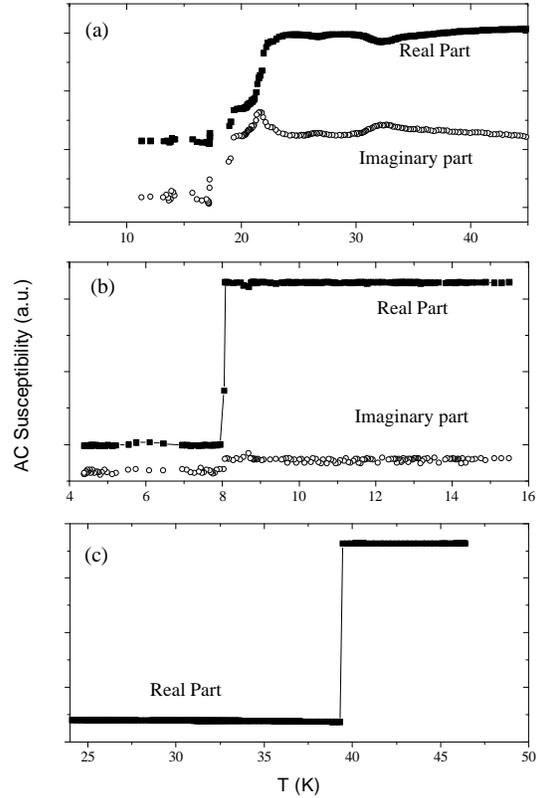

Fig. 3: AC susceptibility of MgB$_2$ films prepared by *in-situ* process (a) with multilayer deposition process, and (b) a single layer film from MgB$_2$ target. For comparison similar measurement data for the film of Fig. 2(a) is shown here in (c).

Fig.2: Magnetization of MgB$_2$ films prepared by *ex-situ* process starting from (a) boron film on STO(100), (b) film deposited from MgB$_2$ target on STO(100), and (c) film deposited from MgB$_2$ target STO(111). For all samples, the applied field was perpendicular to the plane of the film.